\documentclass[12pt]{amsart}
\usepackage{graphicx} 
\usepackage[ruled,vlined]{algorithm2e}
\usepackage[foot]{amsaddr}
\usepackage{amsmath}
\usepackage{amssymb}
\usepackage{amsthm}
\usepackage{array}
\usepackage[authoryear]{natbib}
\usepackage{bigints}
\usepackage{booktabs}
\usepackage{caption}
\usepackage{graphicx, hyperref}
\usepackage{indentfirst}
\usepackage{mathptmx}
\usepackage{mathtools}
\usepackage{mathrsfs}
\usepackage{multirow}
\usepackage{xcolor}
\usepackage[nice]{nicefrac}
\usepackage[paperwidth=8.5in, paperheight=11in, margin=0.5in,headheight=64pt,marginparwidth=0pt,textheight=650pt]{geometry}
\usepackage{tikz}
\usetikzlibrary{arrows.meta, positioning, shapes.geometric}

\title[Bayesian Nonparametric Classifier]{From Partial Exchangeability to Predictive Probability: A Bayesian Perspective on Classification}
\author{Marcio A. Diniz}\thanks{Department of Statistics, Federal University of São Carlos, Rod. Washington Luis, km 235, São Carlos, Brazil. Email: marciodiniz@ufscar.br. Corresponding author.}
\date{}

\begin{document}
\maketitle

\begin{abstract}
We propose a novel Bayesian nonparametric classification model that combines a Gaussian process prior for the latent function with a Dirichlet process prior for the link function, extending the interpretative framework of de Finetti's representation theorem and Ferguson's construction for random distribution functions. This approach allows for flexible uncertainty modeling in both the latent score and the mapping to probabilities.
We demonstrate the method’s performance using simulated data where it outperforms standard logistic regression.
\end{abstract}

\section{Introduction}

Can we model class probabilities without assuming a fixed link function?
Bayesian models for binary classification often assume fixed parametric links. But what if we learn the link itself?
Classical approaches such as logistic regression model the probability of a binary response as a deterministic function of a linear combination of covariates. More flexible methods, such as Gaussian Process Classification (GPC), replace the linear predictor with a latent function drawn from a Gaussian Process, while retaining a fixed link function (typically the logistic or probit CDF).

While these approaches are effective in many applications, they impose strong assumptions on the functional form of the link between the latent score and the class probability. In particular, the use of a fixed link function limits the model’s ability to capture data-driven deviations from standard sigmoidal behavior. Moreover, the probabilistic interpretation of such models often lacks an explicit foundation in the principles of Bayesian inference under exchangeability.

In this paper, we propose a Bayesian nonparametric model for classification that generalizes both logistic regression and GPC. Our approach is motivated by de Finetti’s concept of partial exchangeability: when outcomes depend on covariates, exchangeability does not hold globally, but may still apply conditionally on a latent function. We assume that conditional on a latent function $f$, the labels $Y_i$ are independent, and their probability is determined through a random cumulative distribution function $G$, leading to the hierarchical structure:
\[
f \sim GP(m, K), \quad G \sim DP(\alpha, G_0), \quad P(Y_i = 1 \mid x_i) = G(f(x_i)),
\]

\noindent
where $GP$ denotes a Gaussian process with mean function $m$ and covariance function $K$ and $DP$ a Dirichlet process with concentration parameter $\alpha$ and base measure $G_0$.

This construction leverages two classical results in Bayesian theory: the use of Gaussian processes for modeling latent functions, and Ferguson’s (1973) closed-form posterior for the Dirichlet Process prior on unknown distribution functions. The resulting model is flexible, fully probabilistic, and admits a tractable inference scheme combining MCMC sampling of the latent function with closed-form updates for the link.

We demonstrate the performance of the proposed model through synthetic experiments, comparing its predictive accuracy and calibration with logistic regression. The results show that the model can adapt to nonlinear and non-logistic patterns in the data, while retaining interpretability and providing credible intervals for class probabilities.

The remainder of the paper is organized as follows: Section 2 reviews the theoretical motivation, including de Finetti’s and Ferguson’s results. Section 3 presents the proposed model, and Section 4 details the inference strategy. Section 5 reports simulation results, and Section 6 concludes with a discussion of the method’s limitations and possible extensions.

\section{Theoretical Motivation}

A probabilistic foundation for regression and classification models is given by the notion of \emph{partial exchangeability}, first introduced by \cite{finetti1938} for binary random variables. He illustrated this concept with a simple example: tossing two coins that look similar but may not be identical, leading to exchangeability within each coin’s tosses but not necessarily across them. This motivates grouping observations into classes within which exchangeability holds.

Formally, a sequence of binary random variables divided into $m$ groups is partially exchangeable if the joint distribution is invariant under permutations within each group. The representation theorem states that for such a sequence there exists a unique probability measure $\mu$ on $[0,1]^m$ such that:

\[
P\Big(\sum_{i=1}^{n_1} Y_i(g_1)=y_{1}, \ldots, \sum_{i=1}^{n_m} Y_i(g_m)=y_{m}\Big) = \int_{[0,1]^m} \prod_{j=1}^m \binom{n_j}{y_j} \theta_{j}^{y_j} (1-\theta_j)^{n_j-y_j} d\mu(\theta),
\]

\noindent
where $Y_i(g_j)$ represents the $i$-th binary outcome of the $j$-th group, $i=1, \ldots n_j$ and $j=1, \ldots, m$.

In practice, covariates break full exchangeability, but partial exchangeability still applies: given covariates and a latent function, outcomes are independent. This underlies classical models like logistic regression, where

\[
P(Y=1 | X=x) = G(f(x)),
\]

\noindent
with $f(x)$ often linear in $x$ and $G$ a fixed link function, such as the logistic or Gaussian CDF.

\citet[Section 4.6.4]{bernardo} discuss this framework in dose-response models: animals receive doses and survival is modeled by a logistic link with uncertain parameters. This is a Bayesian logistic regression where $G$ is fixed but parameters are random, and reparameterizations like LD50 make the model interpretable.

Gaussian process classification (GPC) generalizes $f$ to follow a Gaussian process prior, capturing non-linear effects but still assumes a fixed $G$. Our proposal extends this by modeling $G$ itself as unknown, with a Dirichlet process prior, 

\[
P(Y=1 | X=x) = G(f(x)), \quad G \sim DP(\alpha, G_0).
\]

\noindent
where $\alpha$ is a positive scalar and $G_0$ a base probability measure on $(\mathscr{X}, \mathscr{A})$.\footnote{Let $\mathscr{X}$ be a space, usually the sample space, $\mathscr{A}$ a $\sigma$-field of subsets of this space and $G_0$ a finite non-null measure on $(\mathscr{X}, \mathscr{A)}$, in most applications, $(\mathbb{R}, \mathscr{B})$, $\mathscr{B}$ the $\sigma$-filed of Borel sets.}
This keeps partial exchangeability while allowing flexible learning of the link function without imposing a fixed parametric form. 
Ferguson’s classic result shows the DP provides a nonparametric prior for cumulative distribution functions.
Informally, the latent GP transforms the input space into a score space, where the DP estimates the link as a random CDF conditional on these scores.
Therefore, the proposed model can be seen as an explicit implementation of de Finetti’s partial exchangeability for binary regression with covariates, where both the latent structure and the link remain nonparametric.

\section{Proposed Model}


\cite{ferguson1973} describes a typical nonparametric statistical decision problem as follows: the parameter space is the set of all probability measures $G$ on $(\mathscr{X}, \mathscr{A})$. 
The statistician should choose an action $a$ in some space, incurring in a loss $L(G,a)$, basing her choice on an available sample $\mathbf{z}=(z_1, \ldots, z_n)$ from $G$.
Under a quadratic loss function, the Bayes rule with respect to the prior distribution, $G \in DP(\alpha, G_0)$, is, for any $t$ and $F(t)=G((-\infty,t])$ the distribution function to be estimated:

\[
\widehat{F}_B(t \mid \mathbf{z}) 
= \frac{ \alpha G_0((-\infty, t]) + \sum_{i=1}^n \delta_{z_i}((-\infty, t]) }{ \alpha G_0(\mathbb{R}) + n }
= \gamma_n F_0(t) + (1 - \gamma_n) F_n(t \mid \mathbf{z})
\]

\noindent
where

\[
\gamma_n = \frac{ \alpha G_0(\mathbb{R}) }{ \alpha G_0(\mathbb{R}) + n } \qquad \text{and} \qquad F_0(t) = \frac{G_0((-\infty, t])}{G_0(\mathbb{R})}
\]

\noindent
represents the prior guess at the shape of the unknown $F(t)$,  and

\[
F_n(t \mid \mathbf{z}) 
= \frac{1}{n} \sum_{i=1}^n \delta_{z_i}((-\infty, t])
\]

\noindent
the empirical distribution function of the sample.

In our model, Ferguson’s observed sample is replaced by  $(\mathbf{x, \mathbf{y)=}}\{ (x_i, y_i) \}_{i=1}^n$, where $y_i$ is the target (binary) variable and $x_i$ a covariate vector associated with individual $i$.
Therefore we may use $\{ f(x_i)\}_{i=1}^n$—the covariates mapped through the latent function $f$—as a pseudo-sample for the DP.  
Thus, instead of direct draws from the population, we observe latent GP-transformed points and, by Ferguson's results, the DP then estimates the link’s CDF given these points.


This double nonparametric hierarchy can be seen as an operational version of partial exchangeability with unknown conditional CDF, in line with de Finetti’s original insight.
More formally, the priors for the latent and link functions are, respectively:

\[
\mathbf{f} \sim GP(m, K), \qquad G \sim DP(\alpha, G_0)\]

\noindent
where $\mathbf{f}=(f(x_1), \ldots, f(x_n))$, and the likelihood follows from the assumption that the $Y_i's$ are conditionally i.i.d. Bernoulli($G(f(x_i))$, i.e.

\[
P(Y_i = 1 \mid x_i, f, G) = G(f(x_i)).
\]

Ferguson's result mentioned above is used, allowing the posterior of the link function $G$ to be written in closed-form as:

\[
G(f(x_i)) 
= \frac{ 
  \alpha G_0((-\infty, f(x_i)]) 
  + \sum_{j=1}^n \delta_{f(x_j)}((-\infty, f(x_i)]) 
}{
  \alpha G_0(\mathbb{R}) + n 
}.
\]

Given observations $(x_i, y_i)$, the joint posterior for $(\mathbf{f}, G)$ is proportional to:

\begin{equation}\label{eq:post}
p_n(\mathbf{f}, G \mid \mathbf{x}, \mathbf{y})
\propto 
\prod_{i=1}^n 
\big[ G(f(x_i)) \big]^{y_i}
\big[ 1 - G(f(x_i)) \big]^{1 - y_i}
\times 
N_n(\mathbf{f}\mid m, K).    
\end{equation}

\noindent
where $N_n$ denotes a multivariate Gaussian of dimension $n$.
In our experiments, we set the concentration parameter $\alpha = 1$ and use the logistic distribution as the base measure.

It is worth noting that the standard Gaussian Process Classification (GPC) model is recovered as a special case when the link function $G$ is fixed and known (e.g., the logistic or probit CDF). In this sense, the proposed approach can be viewed as a strict Bayesian generalization of GPC, allowing the data to inform the shape of the link via a random distribution function.

\section{Inference Strategy}

Direct posterior sampling from the full posterior, equation \eqref{eq:post}, poses a significant computational challenge: 
it requires a Markov Chain Monte Carlo (MCMC) algorithm that alternates between sampling the latent function $\mathbf{f}$ and the random link function $G$. 
While the posterior for $G$ given $\mathbf{f}$ admits a closed-form (Ferguson's Beta representation), sampling $\mathbf{f}$ conditional on $G$ no longer yields a closed-form marginal likelihood. 
This breaks the standard conjugacy and complicates the use of gradient-based Hamiltonian Monte Carlo (HMC).

To address this, we adopt a practical approximation: 
we first sample $\mathbf{f}$ using a standard GP classification model with a logistic link. 
This step assumes the likelihood
\[
y_i \mid f(x_i) \sim \text{Bernoulli}(\text{logit}^{-1}(f(x_i))),
\]
which provides an efficient, gradient-based inference scheme using HMC. 
The logistic link acts as a convenient parametric surrogate for the true, unknown link.

After obtaining posterior samples for $\mathbf{f}$, we discard the logistic form at the prediction stage and instead estimate the probability $P(Y=1 \mid f(x^*))$ using Ferguson's closed-form DP posterior:
\[
G(f(x^*)) \mid \mathbf{f} 
\sim \text{Beta}\big(
  \alpha G_0(f(x^*)) + m, \;
  \alpha (1 - G_0(f(x^*))) + n - m
\big),
\quad m = \#\{ f(x_j) \le f(x^*) \}.
\]

This hybrid scheme decouples the computational steps: 
it leverages the parametric logistic link for efficient posterior sampling of the latent function while recovering a flexible, nonparametric link via the Dirichlet Process for prediction. 
In practice, this provides a robust approximation to the full model while maintaining computational tractability.

It is worth emphasizing that this two-step approximation has a solid epistemological foundation: the Dirichlet Process posterior for the link function $G$ depends on the observed labels $\mathbf{y}$ only through the latent function $\mathbf{f}$. Formally, conditional on $\mathbf{f}$, the posterior factorizes as
\[
p_n(G \mid \mathbf{f}, \mathbf{x}, \mathbf{y}) = p_n(G \mid \mathbf{f}, \mathbf{x}),
\]
meaning that all information from the data $\mathbf{y}$ is funneled through $\mathbf{f}$. This justifies treating the posterior of $G$ as conditionally independent of $\mathbf{y}$ once the latent structure has been estimated, and allows for tractable inference via Ferguson’s representation.

\section{Simulation Study}

To evaluate the predictive performance and uncertainty quantification of the proposed DP+GP classifier, we conducted two synthetic experiments based on nonlinear classification tasks: \texttt{make\_moons} and \texttt{make\_circles} from \texttt{scikit-learn}.\footnote{Code and simulation notebooks are available at \href{https://github.com/xxxxx/DP-GP}{github.com/yourusername/DP-GP}.}

These datasets are designed to challenge linear models and highlight the benefit of flexible, nonparametric approaches.

\subsection{Experimental setup 1}

We generated a sample of $n=1000$ observations using the \texttt{make\_moons} function from \texttt{scikit-learn}, with added Gaussian noise ($\sigma=0.3$). The sample was randomly split into a training set ($70\%$) and a test set ($30\%$). Each data point $(x_i, y_i)$ consists of a two-dimensional input $x_i \in \mathbb{R}^2$ and a binary label $y_i \in \{0,1\}$.
We compared two models:

\begin{itemize}
    \item \textbf{Logistic regression (LogReg):} A standard parametric logistic regression with linear predictor.
    \item \textbf{DP+GP model:} The proposed Bayesian nonparametric model combining a Gaussian Process prior for the latent score function $f(x)$ and a Dirichlet Process prior for the link function $G$.
\end{itemize}

For the DP+GP model, we implemented inference using NumPyro with the following settings:
\begin{itemize}
    \item \textbf{Latent function:} $f \sim GP(0, K)$ with squared exponential kernel:
    \[
    K(x, x') = \sigma^2 \exp\left( -\frac{\|x - x'\|^2}{2\ell^2} \right) + \sigma_\epsilon^2 \delta(x,x').
    \]
    \item \textbf{Link function:} $G \sim \text{DP}(\alpha, G_0)$, where $\alpha=1$ and $G_0$ is the logistic CDF:
    \[
    G_0(u) = \frac{1}{1 + \exp(-u)}.
    \]
\end{itemize}

Posterior inference on $f$ was performed via HMC using 1000 warm-up steps and 1000 posterior samples. For each test point $x^*$, we drew posterior samples of $f(x^*)$, and evaluated the posterior predictive probability $P(Y^*=1 \mid x^*)$ via Ferguson’s Beta posterior for $G(f(x^*))$. For each point, we also computed $95\%$ credible intervals based on the quantiles of the sampled values.


Figure~\ref{fig:dp-gp-boundary} shows the predictive decision boundary for the DP+GP model, including credible regions (2.5\% and 97.5\% quantiles). The decision boundary adapts flexibly to the nonlinear geometry of the data, and the credible bands reflect greater uncertainty near the boundary and in regions with sparse training data.

\vspace{1em}
\begin{center}
\includegraphics[width=0.7\textwidth]{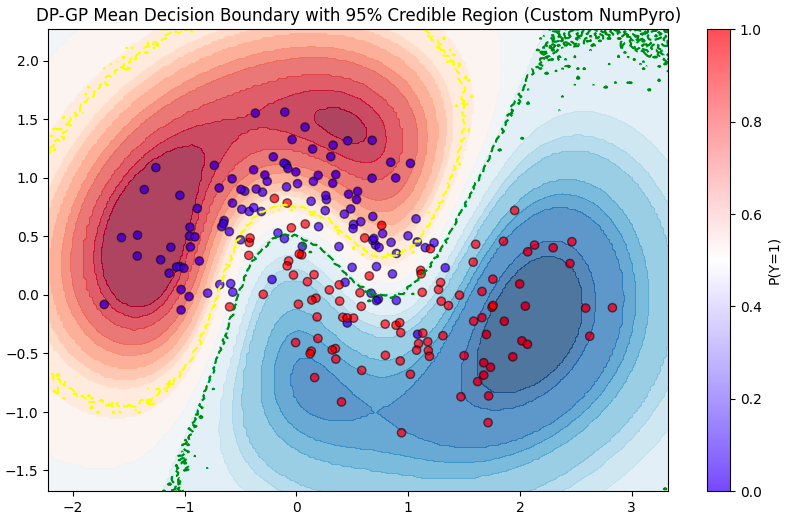} 
\captionof{figure}{Predictive decision boundary of the DP+GP model with 95\% credible bands.}
\label{fig:dp-gp-boundary}
\end{center}
\vspace{1em}

We evaluated the models on the test set using three standard classification metrics:
\begin{itemize}
    \item Area Under the ROC Curve (AUC),
    \item Brier Score,
    \item Logarithmic Loss (LogLoss).
\end{itemize}

The results are summarized below in Table \ref{tab:moons}.

\begin{table}[h!]
\centering
\caption{Performance comparison on \texttt{make\_moons} dataset}
\label{tab:moons}
\begin{center}
\begin{tabular}{lccc}
\toprule
\textbf{Model} & \textbf{AUC} & \textbf{Brier Score} & \textbf{LogLoss} \\
\midrule
DP+GP & 0.964 & 0.099 & 0.338 \\
Logistic Regression & 0.939 & 0.096 & 0.311 \\
\bottomrule
\end{tabular}
\end{center}
\end{table}

The DP+GP model outperforms logistic regression in terms of AUC, indicating better discriminative ability, while maintaining comparable Brier and LogLoss scores, reflecting good calibration. 

\subsection{Experimental setup 2}

The second experiment used the \texttt{make\_circles} function, which produces data lying on two concentric rings—representing a challenging task for linear classifiers. As before, we generated $n = 1000$ samples with Gaussian noise ($\sigma = 0.1$), and split the dataset into 70\% training and 30\% test sets.

For computational efficiency, we used a lightweight version of the DP+GP model: we fixed GP kernel hyperparameters and computed the posterior mean function analytically. Ferguson’s closed-form update was then used to compute $P(Y = 1 \mid x^*)$ and its 90\% credible interval.

\begin{figure}[h!]
\centering
\includegraphics[width=0.7\textwidth]{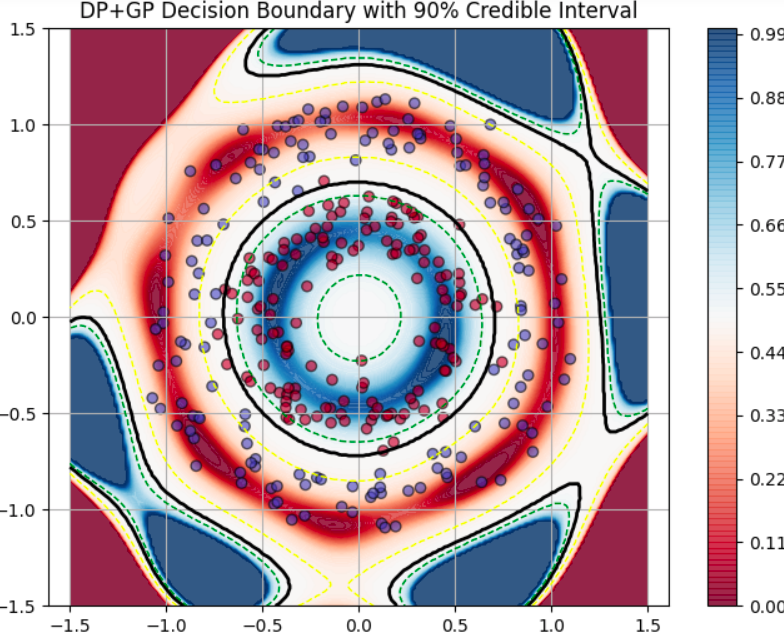}
\caption{
Decision boundary of the DP+GP classifier on the \texttt{make\_circles} dataset. The black line corresponds to the predictive contour $\mathbb{E}[G(f(x))] = 0.5$. Dashed green and yellow lines indicate the 90\% credible region around the decision boundary, computed via Ferguson’s posterior. The model accurately recovers the circular structure and quantifies predictive uncertainty.
}
\label{fig:circles}
\end{figure}

Figure~\ref{fig:circles} shows the predicted decision boundary (solid black line) and credible region (dashed yellow and green lines). Table~\ref{tab:circles} reports the predictive performance.

\begin{table}[h!]
\centering
\caption{Performance comparison on \texttt{make\_circles} dataset}
\label{tab:circles}
\begin{center}
\begin{tabular}{lccc}
\toprule
\textbf{Model} & \textbf{AUC} & \textbf{Brier Score} & \textbf{LogLoss} \\
\midrule
DP+GP & 0.9997 & 0.103 & 0.358 \\
Logistic Regression & 0.473 & 0.251 & 0.696 \\
\bottomrule
\end{tabular}
\end{center}
\end{table}

\section{Conclusions}

This paper proposed a novel Bayesian nonparametric model for binary classification that combines a Gaussian Process (GP) prior over a latent score function $f$ with a Dirichlet Process (DP) prior over the link function $G$. The model generalizes logistic regression and standard GP classification by replacing the fixed logistic link with a random cumulative distribution function, enabling greater flexibility and epistemic uncertainty modeling.

Our formulation draws on the foundational results of de Finetti and Ferguson. Partial exchangeability motivates the latent functional structure, while Ferguson’s posterior representation of the DP allows for tractable, closed-form inference on the link function. Empirical experiments show that the proposed DP+GP model performs competitively with logistic regression, especially in nonlinear classification tasks.

\noindent
\textbf{Limitations and future work.} While the model is appealing from both a theoretical and practical standpoint, it also presents some limitations:

\begin{itemize}
  \item \textit{Computational complexity.} The approach is fully Bayesian only conditional on fixed hyperparameters. A fully hierarchical nonparametric model would require priors on all hyperparameters (e.g., $\alpha$, kernel lengthscale $\ell$, base measure $G_0$), resulting in slower MCMC sampling and greater computational cost.
  
  \item \textit{Discrete nature of the DP.} The Dirichlet Process yields discrete random measures, which can introduce staircase-like artifacts in the estimated link function, particularly when smoother transitions are desirable.
  
  \item \textit{Efficiency in linear settings.} When the relationship between covariates and outcomes is approximately linear, classical parametric models such as logistic regression remain more computationally efficient and often comparably accurate.
\end{itemize}

Future work could explore variational inference techniques to improve scalability, investigate alternative priors for smoother link functions (e.g., logistic Gaussian Processes or mixtures of CDFs), and apply this methodology to real-world datasets where covariate shifts and complex noise structures are present.

Another direction points to practical limitations regarding the concept of partial exchangeability.
While partial exchangeability offers a compelling probabilistic foundation for modeling covariate-dependent data, it has important limitations in practice. Formally, the assumption asserts invariance under permutations of observations that share the same covariate values, leading to a mixture representation with group-specific random measures. However, this assumption breaks down in typical applications where covariates are continuous and repeated values are rare or absent.

In such settings, partial exchangeability does not enforce the desirable property that observations with similar covariates should have similar distributions. For instance, permuting two observations with covariates $x$ and $x'$ close in Euclidean distance may lead to substantial changes in the likelihood, violating intuitive notions of “smoothness” or “local coherence.” This is particularly problematic when attempting to regularize or constrain the space of latent mixing distributions $(G_x)_{x \in \mathcal{X}}$, since partial exchangeability allows arbitrary dependence across covariates.

To address this gap, recent work introduces the notion of \emph{local exchangeability} \citep{campbell2022local}, which relaxes the invariance requirement by bounding the total variation distance between distributions under local permutations. This framework provides a principled middle ground between full exchangeability and rigid conditional independence, and yields a generalized de Finetti-type representation in terms of a continuous measure-valued stochastic process. Incorporating such a notion could strengthen the theoretical foundations of models like the one proposed in this paper, especially in high-dimensional or densely sampled covariate spaces.

This work illustrates how classical Bayesian concepts can be operationalized into modern, flexible tools for classification, with tractable inference and principled uncertainty quantification.

\appendix
\section{Two-step inference strategy for the DP+GP model}

We describe here the posterior inference procedure for the proposed model combining a Gaussian Process prior for the latent function $f$ and a Dirichlet Process prior for the link function $G$.

\noindent
\textbf{Model:}
\begin{align*}
\mathbf{f} &\sim GP(m, K), \\
G &\sim DP(\alpha, G_0), \\
Y_i \mid f, G &\sim \text{Bernoulli}(G(f(x_i))).
\end{align*}

\noindent
\textbf{Inference Pseudocode:}

\begin{algorithm}[H]
\caption{Two-step inference strategy for the DP+GP model}
\KwIn{Training data $\{(x_i, y_i)\}_{i=1}^n$, kernel $K$, base measure $G_0$, concentration $\alpha$}
\KwOut{Posterior predictive distribution for test points $\{x_j^*\}_{j=1}^m$}

\textbf{Step 1: Posterior over latent function $f$}\;
Sample $\mathbf{f} \sim GP(m, K)$ via HMC using a logistic GP auxiliary model\;
  
\textbf{Step 2: Posterior over $G$ via Ferguson's Beta representation}\;
\ForEach{posterior sample $\mathbf{f}^{(s)}$}{
    \ForEach{test point $x_j^*$}{
        Compute $f^{(s)}(x_j^*)$ via GP posterior\;
        Compute $m^{(s)} = \# \{ f^{(s)}(x_i) \le f^{(s)}(x_j^*) \}$\;
        Sample $G(f^{(s)}(x_j^*)) \sim \text{Beta}( \alpha G_0(f) + m,\; \alpha (1 - G_0(f)) + n - m )$\;
        Set $P^{(s)}(Y=1 \mid x_j^*) = G(f^{(s)}(x_j^*))$\;
    }
}
\textbf{Step 3: Aggregate predictions}\;
Compute predictive mean and quantiles over $s=1,\dots,S$ for each $x_j^*$\;
\end{algorithm}

\newpage

\section*{Data Availability}


The code used to generate the examples discussed above are available upon request.\footnote{Author's email: marciodiniz@ufscar.br.}

\section*{Acknowledgments}

During the preparation of this manuscript, the author used ChatGPT to assist in developing the simulation code. All content generated with the help of this tool was subsequently reviewed and edited by the author, who takes full responsibility for the final version of the work.


\begin{thebibliography}{4}
\providecommand{\natexlab}[1]{#1}
\providecommand{\url}[1]{\texttt{#1}}
\expandafter\ifx\csname urlstyle\endcsname\relax
  \providecommand{\doi}[1]{doi: #1}\else
  \providecommand{\doi}{doi: \begingroup \urlstyle{rm}\Url}\fi

\bibitem[Bernardo and Smith(1994)]{bernardo}
J.~M. Bernardo and A.~F. Smith.
\newblock \emph{Bayesian Theory}.
\newblock Wiley, 1994.

\bibitem[Campbell et~al.(2022)Campbell, Syed, Yang, Jordan, and Broderick]{campbell2022local}
Trevor Campbell, Saifuddin Syed, Chiao-Yu Yang, Michael~I Jordan, and Tamara Broderick.
\newblock Local exchangeability.
\newblock \emph{arXiv preprint arXiv:1906.09507}, 2022.

\bibitem[de~Finetti(1938)]{finetti1938}
B.~de~Finetti.
\newblock Sur la condition d'échangeabilité partielle.
\newblock \emph{Annales de l'I.H.P.}, 10:\penalty0 119--125, 1938.

\bibitem[Ferguson(1973)]{ferguson1973}
Thomas~S Ferguson.
\newblock A {B}ayesian analysis of some nonparametric problems.
\newblock \emph{The Annals of Statistics}, 1\penalty0 (2):\penalty0 209--230, 1973.
\newblock \doi{10.1214/aos/1176342360}.

\end{thebibliography}
\end{document}